\documentclass[12pt,preprint]{aastex} 
\bibpunct{(}{)}{;}{a}{}{,} 
 
\newcommand{\n}{\newcommand} 
\n{\bmain}{\begin{list}{${\bullet}$}{}} 
\n{\emain}{\end{list}} 
\n{\bminor}{\begin{list}{$\bf\triangleright$}{}} 
\n{\eminor}{\end{list}} 
\n{\Ms}{\mbox{$\,M_\odot$}} 
\slugcomment{Accepted by Astrophysical Journal Letters} 
 
\shorttitle{The Rates of Hypernovae and GRBs} 
\shortauthors{Podsiadlowski et al.} 
 
\begin{document} 
 
\title{The Rates of Hypernovae and Gamma-Ray Bursts:  
Implications for their Progenitors} 
 
 
\author{Ph.\ Podsiadlowski\altaffilmark{1}, P. A. Mazzali\altaffilmark{2,3,4},  
K. Nomoto\altaffilmark{3}, D. Lazzati\altaffilmark{5},  
E. Cappellaro\altaffilmark{6}} 
 
\altaffiltext{1}{Department of Astrophysics, University of Oxford, 
Oxford, OX1 3RH, UK; podsi@astro.ox.ac.uk } 
 
\altaffiltext{2}{INAF-Osservatorio Astronomico di Trieste, 34131
Trieste, Italy; mazzali@ts.astro.it }
 

\altaffiltext{3}{Dept.\ of Astronomy and RESCEU, University of Tokyo,
Tokyo 113-0033, Japan; nomoto@astron.s.u-tokyo.ac.jp }
 
\altaffiltext{4}{MPI f\"ur Astrophysik, 85748 Garching, Germany} 
 
\altaffiltext{5}{Institute of Astronomy, University of Cambridge,  
Madingley Road, Cambridge CB3 0HA, UK; lazzati@ast.cam.ac.uk } 
 
\altaffiltext{6}{INAF-Osservatorio Astronomico di Capodimonte, 
80131 Napoli, Italy; cappellaro@na.astro.it }



 
\begin{abstract} 
A critical comparison of estimates for the rates of hypernovae (HNe)  
and gamma-ray bursts (GRBs) is presented. Within the substantial  
uncertainties, the estimates are shown to be quite comparable and give 
a Galactic rate of $10^{-6}$\,--\,$10^{-5}\,$yr$^{-1}$ for both events.  
These rates are several orders of magnitude lower than the rate of  
core-collapse supernovae, suggesting that the evolution leading to a HN/GRB  
requires special circumstances, very likely due to binary interactions.  
Various possible binary channels are discussed, and it is shown that these  
are generally compatible with the inferred rates. 
\end{abstract}



\section{Introduction} 
 
While it has now been established for more than 5 years that gamma-ray 
bursts (GRBs) are caused by some of the most energetic explosions in 
the Universe (van Paradijs, Kouveliotou, \& Wijers 2000), no promising 
channel for their progenitors has been identified, a situation very 
much resembling that of the progenitors of Type Ia supernovae some 20 
years ago. The firm and, unlike the previous case, unambiguous 
association of a GRB (GRB030329) with a hypernova, SN~2003dh (Hjorth 
et al.\ 2003a; Stanek et al.\ 2003), a highly energetic Type Ic 
supernova (Mazzali et al.\ 2003)\footnote{The term Hypernovae has been 
used for SNe with energies significantly larger (by about a factor of 
10 or more) than the canonical explosion energy of $1\,{\rm foe} \equiv 
10^{51}\,$ergs (Nomoto et al.\ 2003).}, has confirmed that at least 
some long-duration GRBs are {\em observationally} connected with the 
explosion of massive stars\footnote{GRBs fall into two classes: short- 
and long-duration bursts. Presently, very little is known about the 
progenitors of short bursts from an observational point of view.  
It is quite possible that they are caused by a completely different 
physical mechanism, e.g., the merger of two compact objects 
(van Paradijs et al.\ 2000). All inferences made in this paper 
exclusively apply to long bursts.}. 
 
All hypernovae known to date belong to the class of Type Ic supernovae
(SNe~Ic), of which they form a subset.  These are SNe that show
neither hydrogen nor significant amounts of helium in their
spectra. Their progenitors are believed to be either very massive
single stars that lost their hydrogen and helium envelopes in a
stellar wind or massive stars that lost their envelopes through the
interaction with a companion (Wheeler \& Levreault 1985; Uomoto 1986;
Podsiadlowski, Joss, \& Hsu 1992; Nomoto et al.\ 1994).  Two out of
the three nearby GRBs known to date are associated with hypernovae
(GRB980425/SN~1998bw at $z = 0.008$ and GRB030329/SN~2003dh at $z =
0.17$). The third case, GRB031203, is heavily extinguished by dust, so
a SN association cannot be firmly ruled out (Hjorth et al.\ 2003b).
This interesting coincidence raises the important question of the
general connection between hypernovae and long-duration GRBs, and
whether most, or perhaps even all GRBs are associated with/caused by
hypernovae.
 
This paper addresses this question by providing a critical comparison 
of the rates of GRBs and hypernovae. As is shown in \S~2, the rates 
of GRBs and hypernovae are comparable within the uncertainties, and 
appear to be a small fraction of the global SN rate. This has 
important implications for the nature of their progenitors, 
which is discussed in detail in \S~3 and \S~4.

\section{The Rates of GRBs and Hypernovae} 
 
\subsection{The GRB rate} 
 
The rate of {\em observed} GRBs in a galaxy like our own is quite well
established from the BATSE monitoring as $R_{\rm obs}\sim
10^{-7}\,$yr$^{-1}$ (e.g. Zhang \& Meszaros 2003).  However, since GRB
fireballs are highly beamed, both geometrically and relativistically
(with Lorentz factors $\Gamma\gtrsim100$), the true intrinsic rate
must be substantially higher. It may be written as $R_{\rm GRB} =
R_{\rm obs}\times {4\pi\over \Omega}$, where $\Omega$ is the solid
angle within which an observer can detect the GRB.  This factor
depends on the jet opening angle, and is typically estimated as $\sim
50$\,--\,500 (Frail et al. 2001; Panaitescu \& Kumar 2001).  The rate
is however uncertain, as $R_{\rm obs}$ and $\Omega$ are estimated from
two different samples, $R_{\rm obs}$ from the BATSE sample, and
$\Omega$ from the sample of GRBs with afterglow observations, and
there is no robust evidence that the selection effects are the same in
the two sets.  Additionally, the solid angle correction is based on
the so-called uniform jet model, in which the opening angle is an
intrinsic property of the jet. An alternative explanation of the
observations calls for a structured jet, with a brighter core and
dimmer wings. In this case the rate of GRBs would be smaller by a
factor 3\,--\,10 (Rossi, Lazzati, \& Rees 2002). Thus the range of
plausible values for the GRB rate is $10^{-6}$ to $10^{-5}\,$yr$^{-1}$, 
of which about $2/3$ are long-duration GRBs.

\subsection{The Hypernova rate} 
 
To date, five SNe Ic have been classified as hypernovae. 
They form quite a diverse group of objects, ranging from the very bright 
and energetic SNe 1998bw (Iwamoto et al.\ 1998) and 2003dh (Mazzali et al.\ 
2003), the moderately bright but very energetic SNe 1997ef (Mazzali et al.\ 
2000) and 1997dq (Mazzali et al., in preparation), and the normally bright 
but over-energetic SN~2002ap (Mazzali et al.\ 2002). Based on spherically 
symmetric explosion models, their explosion energies have been estimated 
to range between 4 and 50\,foe, and the progenitor masses from a lower 
limit of 20\,--\,25\Ms\ to 40\Ms\ and above. This covers the entire 
mass range of single stars that are believed to become black holes 
(e.g. Maeder 1992; Fryer \& Kalogera 2001). 

Interestingly, no hypernova is known to have the characteristics of a
SN~II, although such objects might in principle occur in the lower
mass range (depending somewhat on the minimum initial mass for which a
{\em single} star becomes a Wolf-Rayet star). This may provide an
important clue, linking the physical cause of the hypernova mechanism
to the process causing the loss of the hydrogen and helium envelope.
In this context, we note as a caveat that the inferred initial masses
of hypernovae are based on the final core structure expected from
single-star evolution. If, the pre-hypernova  evolution was affected
by binary evolution, as seems possible or even likely (see \S~3), this
mapping must be modified \footnote{For example, Brown et al.\ (2001)
showed that, if a star loses its envelope through a binary interaction
soon after its main-sequence phase, its final pre-supernova core
structure is dramatically changed and even a 60\,\Ms\ star may produce
a neutron star rather than a black hole (also see Podsiadlowski
et al.\ 2003).}.

As the lowest initial mass that is able to produce a hypernova appears
to be $\sim 20\Ms$, this implies that not all SNe Ic are hypernovae.
For example, Nomoto et al.\ (1994) estimate that the progenitor of the
normal SN Ic 1994I was 15\Ms\ (again assuming a single-star mapping of
the initial to the final mass).  SNe Ic may come from progenitors as
low in mass as $\sim 8\Ms$ if they are in a binary (Podsiadlowski et
al.\ 1992; Nomoto et al.\ 1994).

The estimated rate of all core-collapse supernovae is
$7\times10^{-3}\,$yr$^{-1}$ for an average galaxy and $1.2\times
10^{-2} \,$yr$^{-1}$ in our Galaxy (Cappellaro et al.\ 1999). The
latter estimate is somewhat lower than recent estimates for the
Galactic pulsar birth rate of $4\times 10^{-2}$\,yr$^{-1}$ based on
the Parkes multi-beam survey (Vranesevic et al.\ 2003). In contrast,
the observed rate of Type Ib and Ic supernovae in an average galaxy in
the local Universe is only $\sim 10^{-3}\,$yr$^{-1}$.

Most Ib/c supernovae actually appear to belong to the Ic sub-type and
only a fraction of about 5\,\% of {\em observed} SNe Ic are
hypernovae.  The brightness of hypernovae is highly diverse, ranging
from normal to about 10 times normal. However, the average of the
known cases is a factor of $\sim 3$\,--\,5 brighter than a typical
SN~Ic.  Therefore, we expect that hypernovae are easier to detect and
hence intrinsically less common relative to normal SNe Ic than the
direct observational estimate. Being on average a factor 4 brighter
implies that in a magnitude-limited search they would be detectable in
a volume larger by a factor of $4^{3/2}=8$. However, because many of
the current SN searches only target selected galaxies, they are also
volume limited. Thus, in a typical SN search the expected number of
SNe grows more or less linearly with SN magnitude (Cappellaro et
al.\ 1993). Reducing the observed rate by the proper factor gives us an
estimate of the true hypernova rate of $\sim 10^{-5}\,$yr$^{-1}$.

\section{The Progenitor Connection} 
 
The estimates of the rates of GRBs and hypernovae in the previous 
section are the same to within the uncertainties (see Table~1), 
although the hypernova rate may be slightly higher. This suggests that 
most hypernovae also appear as GRBs at least from some viewing angle. 
 
This is consistent with the fact that hypernovae are associated with at 
least two out of three nearby GRBs, one of which was typical while the 
other was weak. It is also consistent with the fact that only the most 
powerful hypernovae are seen in association with GRBs.  
Events that appear less powerful may simply be viewed off-axis, 
leading to an underestimation of the kinetic energy and to the 
non-detection of the GRB.  SN~2002ap could be such a case, 
since there is ample evidence that the explosion was aspherical, 
like SN~1998bw (Mazzali et al.\ 2002, Maeda et al.\  2003). 
This may also imply that the beaming correction cannot be too large. 
 
The estimates allow for the possibility that some hypernovae do not 
produce GRBs, as in some popular models (MacFadyen \& Woosley 1999) 
the relativistic jet may not always break through the envelope of the 
progenitor star. This may give rise to a `failed GRB' with an orphan 
afterglow, as has been suggested for SN~2002ap (Totani 2003), or to 
short-duration X-ray flashes (XRFs; Heise et al.\ 2001).  
Possible evidence for this comes from the reported detection of a 
SN-like ``bump'' in the light curve of an X-ray flash (XRF030723, 
Fynbo et al.\ 2004). 
 
SN-like bumps have been detected in the light curves of GRB optical 
afterglows, but only for one such case (GRB021211/SN2002lt, 
$z \sim 1$) is a spectrum available: Della Valle et al.\  (2003) argue 
that it is similar to that of the standard SN~Ic 1994I.  
However, the extracted SN U-band light curve (other bands not being 
available) appears significantly brighter than the U-band light 
curve of SN~1994I, so a hypernova solution for SN~2002lt cannot be 
firmly ruled out. If a clear case of association of a normal SN~Ic 
and a GRB should be revealed, we may have to lower the mass limit 
of stellar collapses that trigger a GRB. 
 
These estimates are also in broad agreement with those of Berger et
al.\ (2003) and Soderberg, Frail, \& Wieringa et al.\ (2003).  Based
on a comparison of the radio emission from hypernovae and SNe Ib/c,
these studies conclude that $\la 3\,$\% and $\la 6\,\%$, respectively,
of SNe Ib/c can be associated with GRBs.  In contrast, Lamb et al.\
(2003) recently argued that, based on a universal jet model for XRFs
and GRBs, the jet opening angle is as small as 0.5$^{\circ}$. This
would imply an XRF/GRB rate comparable to the SN Ib/c rate.
 
Our rate estimates suggest that GRBs and hypernovae constitute a small
subset of core-collapse supernovae (also see, Paczy\'nski 2001).  Does
this imply that only very massive stars become HNe/GRBs?  In
Table~1 we list the estimated rates for stars above various different
masses, using a simple Salpeter-like mass function ($f(M)\, {\rm d}
M\propto M^{-2.5}\,{\rm d} M$) and assuming for simplicity that all
stars above 8\Ms\ produce a core-collapse supernova. Clearly, even if
the minimum initial mass for a HN/GRB was larger than 80\Ms,
they would be significantly overproduced.  On the other hand, the
initial progenitor mass in some hypernovae appears to be as low as 20\Ms\
(Mazzali et al.\ 2002; but see footnote 9).

\begin{deluxetable}{lc}
\tablecolumns{2} 
\tablewidth{\textwidth} 
\tablecaption{Rates in an average galaxy} 
\tabcolsep=4pt 
\tablehead{&Rate (yr$^{-1}$)} 
\startdata 
Core-collapse supernovae& $7\times 10^{-3}$\\ 
Radio pulsars (Galactic)& $4\times 10^{-2}$\\ 
SNe Ib/c& $1\times 10^{-3}$\\ 
Hypernovae& $\sim 10^{-5}$\\ 
\noalign{\vspace{2pt}}
\multicolumn{2}{l}{GRBs (for different effective beaming angles $\theta$)}\\ 
\noalign{\vspace{2pt}}
\hspace{1cm}$\theta=1^\circ$&$6\times 10^{-4}$\\ 
\hspace{1cm}$\theta=5^\circ$&$3\times 10^{-5}$\\ 
\hspace{1cm}$\theta=15^\circ$&$3\times 10^{-6}$\\ 
Massive stars\\ 
\hspace{1cm}$> 20\Ms$&$2\times 10^{-3}$\\ 
\hspace{1cm}$> 40\Ms$&$6\times 10^{-4}$\\ 
\hspace{1cm}$> 80\Ms$&$2\times 10^{-4}$\\ 
\enddata 
\end{deluxetable} 
 
In conclusion, it is extremely unlikely that the progenitors of
hypernovae and GRBs are just very massive stars. Special circumstances
are almost certainly needed. The most promising of these is rotation:
a rapidly rotating core is the essential ingredient in the `collapsar'
model for GRBs (Woosley 1993; Paczy\'nski 1998; MacFadyen \& Woosley
1999). The prototype hypernova, SN 1998bw, shows clear evidence from
the line profiles that the explosion was highly asymmetric (Maeda et
al.\ 2002).

\par\medskip 
\noindent{\em The role of rotation} 
\par\medskip 
\noindent 
Massive stars are generally rapid rotators on the main sequence. 
However, there are many well-established mechanisms by which 
they can lose their angular momentum during their evolution by 
both hydrodynamical (e.g.\ winds) and magnetohydrodynamical 
processes (Spruit \& Phinney 1998; Spruit 2002).  
Therefore, it is not at all clear whether the cores of massive 
single stars will ever be rotating rapidly at the time of explosion. 
In this context, rapid rotation means sufficiently rapid that the 
core cannot collapse directly to form a neutron star/black hole 
and conserve angular momentum. A simple criterion is that the 
specific angular momentum, $j$, near the edge of the iron core 
(enclosing a mass $M_c \sim 2\Ms$) is larger than the value at the 
last stable orbit around a black hole of that mass, i.e.\ 
$j\ga \sqrt{6} GM_c/c\simeq 2\times 10^{16}\mbox{\,ergs\,s}\,(M_c/2\Ms)$. 
Recent calculations taking into account magnetic torques (Heger et al.\ 
2003) suggest that single massive stars fall short of this requirement 
by about one order of magnitude. 
 
To have a sufficiently rapidly rotating core at the time of  
explosion may require interactions with a binary companion that 
can spin up the progenitor or prevent its spin-down. 
 
\par\medskip 
\noindent{\em The role of binarity} 
\par\medskip 
\noindent 
Binary interactions can spin up a star by a variety of processes. 
Tidal interactions can cause either component of a binary to rotate 
with the same frequency as the binary, spinning it up or down 
depending on the relative frequencies. For a star spinning 
synchronously with the binary orbit and filling a fraction $r$ of its 
Roche lobe, the ratio of its rotation frequency, $\omega$, to its 
(Keplerian) breakup frequency, $\omega_{\rm crit}$, just depends on 
the mass ratio according to 
${\omega/\omega_{\rm crit}} = (1+q)^{1/2}\,h(q)^{3/2}\,r^{-3/2},$ 
where $q=M_1/M_2$ is the mass ratio ($M_2$ is the Roche-lobe filling 
object and $M_1$ the accreting star) and $h(q)$ is the ratio of the 
Roche-lobe radius to the orbital separation (as, e.g., given by 
Eggleton 1983). 
 
If we require that in a collapsar model only the inner-most core of
$\sim 2\Ms$ can collapse directly while the rest forms a disk, we can
obtain a rough estimate for the maximum orbital period where tidal
spin-up can provide enough angular momentum to the core by assuming
that the whole star remains in solid body rotation until the end of
helium burning. At this stage the core is likely to decouple and will
probably retain most of its angular momentum in the final rapid
evolutionary phases.  Taking the radius of the 2\Ms\ core as $\sim
8\times 10^9\,$cm (typical for the core of a 30\Ms\ star at the end of
helium burning), one then immediately obtains a critical orbital
period $P_{\rm crit} \sim 5.6\,{\rm hr}\,(R/8\times 10^9\,{\rm
cm})^2\, (j/2\times 10^{16}\,{\rm ergs\, s})^{-1}$.  Izzard et al.\
(2003), using detailed binary population calculations, concluded that
there are enough binaries where tidal locking could account for the
observed rates\footnote{Note, however, that they assumed that it was
sufficient to prevent the whole star (rather than just the core) from
collapsing directly into a black hole, which significantly increases
the critical orbital period and hence the estimated rate for this
channel.}.

The black-hole binary Nova Sco may provide indirect observational
support for this channel.  The companion in Nova Sco is polluted
with heavy elements from the SN that formed the black hole (Israelian
et al.\ 1999).  Podsiadlowski et al.\ (2002) concluded that the
observed abundances are more consistent with a hypernova than with a
normal SN. The birthrate of such black-hole binaries is comparable to
the HN/GRB rate (see, e.g., Lee, Brown, \& Wijers 2002; Podsiadlowski,
Rappaport, \& Han 2003).

Tidal locking is also likely to have spun up the WR companion (van
Kerkwijk et al.\ 1992) in the close X-ray binary Cyg X-3, which has an
orbital period of 4.8\,hr. Thus Cyg X-3 is a potential HN/GRB
progenitor, although the mass of the WR may be too small to lead to
the formation of a black hole. Nevertheless, the birthrate of systems
like Cyg X-3, one of the main channels to produce double-neutron-star
binaries, is comparable to our estimate of the HN/GRB rate (Dewi \&
Pols 2003; Ivanova et al.\ 2003; Kalogera et al.\ 2003)

The most dramatic type of binary interaction involves the complete
merging of two stars or at least of their cores (Fryer \& Woosley
1998; Zhang \& Fryer 1999; Ivanova \& Podsiadlowski 2003; Joss \&
Becker 2003; Nomoto et al.\ 2003). For example, Ivanova \&
Podsiadlowski (2003) found that the core of the progenitor of SN~1987A
was spun up dramatically in the merger, implying a rapidly
rotating core at the time of the explosion. SN~1987A was not a
hypernova, but there is some evidence that the explosion was jet-like
(e.g., Wang et al.\ 2002; also see Joss \& Becker 2003). Fryer \&
Woosley (1998) suggested that the merger rate of black holes and
helium cores inside a common envelope is compatible with the GRB rate.
None of these merger suggestions for HNe/GRBs have yet been worked out
in detail. If the merger is caused by the spiral-in inside a
hydrogen-rich common envelope, one has to understand how the merger
can proceed and still lead to the ejection of the envelope (which is
required to provide the friction for the spiral in).  Moreover, all
hypernovae are SNe~Ic, and thus the progenitors must have lost both
their hydrogen and helium envelopes.

\section{Discussion} 
 
The rates of hypernovae and GRB are quite comparable, suggesting that
a large fraction (most?) of hypernovae also produce GRBs, at least in
some direction.  Moreover, the rates are significantly smaller than
the rates of core-collapse SNe (or even the fraction of SNe that
produce black holes).  Furthermore, at least at the present
cosmological epoch, special circumstances are required to produce
HNe/GRBs.  However, numerous fundamental questions remain unanswered,
and no fully self-consistent evolutionary model for the progenitors
exists at this time.  As long as this is the case, it is not even
clear whether or not the hypernova and the GRB occur concurrently.
Does the hypernova occur first and trigger the GRB through the
fallback of hypernova ejecta, as may be required in some models
(Vietri \& Stella 1998; Podsiadlowski et al.\ 2002), or does the GRB
occur and then trigger a SN-like event through the interaction of the
relativistic jet with the envelope as in the collapsar model (Khokhlov
et al.\ 1999; MacFadyen \& Woosley 1999; Jason, Woosley, \& Hoffman
2003)?
 
Although HNe/GRBs appear to be relatively rare events at the present
epoch, this need not be the case for the first generation of stars.
Lower metallicity may lead to lower angular-momentum loss from
massive stars, and the star-formation environment may be very
different. It is even conceivable that, at an early epoch of galaxy
formation, hypernovae could provide the missing energy to eject 
half the baryons from galaxies (Silk 2003).
 
Finally, another important question concerns the relationship between 
hypernovae and the class of SNe Ib/c, of which they are a subgroup. 
Presumably, many normal SNe Ib/c are caused by the collapse of the 
core of a massive star that lost its H-rich envelope through binary 
interaction (Wheeler \& Levreault 1985; Podsiadlowski et al.\ 1992; 
Nomoto et al.\ 1994) forming a neutron star. So perhaps one important 
distinction between a hypernova and an ordinary supernova is whether a 
black hole or a neutron star is formed in the aftermath. 
However, not all black hole formation events can lead to a hypernova: if 
the minimum mass of a single star that leads to the formation of a black 
hole is as low as 20\,--\,25\Ms\ (Maeder 1992; Fryer \& Kalogera 2001), 
this would overproduce hypernovae by a large factor (see Table 1). 
 
A natural explanation for this dichotomy may lie in the fact that black 
holes can either form promptly on a dynamical timescale or on a much 
longer timescale by continued accretion through a disk phase or fallback.  
In particular, the disk accretion phase, which is the essential 
ingredient in collapsar models, requires a rapidly rotating core. 
In the case of prompt collapse, one would not necessarily expect a 
bright SN.  This would imply the existence of a class of (very?) dim 
SNe Ib/c for which there is, however, no observational evidence at present. 
 
\bigskip
This work was in part supported by a European
Research \& Training Network (HPRN-CT-20002-00303) and a Royal Society
visitor's grant to K.N.
The work of P.P., P.M., and K.N. (the "Tokyo Think Tank") at the University 
of Tokyo has been supported by the 21st Century COE Program of MEXT, Japan.

\clearpage 
 

 
 
\end{document}